\documentclass[prl,aps,showpacs,floatfix,nofootinbib,twocolumn,letterpaper,reprint]{revtex4}
\usepackage{graphicx} 
\usepackage{amssymb} 
\usepackage{amsmath} 
\usepackage{amsfonts}

\def\be{\begin{equation}} 
\def\ee{\end{equation}}

\begin{document} 

\title{
Nuclear fission as resonance-mediated conductance
}
\author{G.F. Bertsch}
 
\affiliation{Department of Physics and Institute of Nuclear Theory, 
Box 351560\\ University of Washington, Seattle, Washington 98915, USA}

\begin{abstract}

Up to now, the theory of nuclear fission has relied on the existence of a
collective coordinate associated with the shape of the nucleus, giving rise
to a spectrum of channels through which the fission takes place. We present
here an alternate formulation of the theory, in which the fission is
facilitated by individual states in the barrier region rather than channels
over the barrier.  In a simplified limit, the theory reduces to a well-known
formula for electronic conductance through resonant tunneling states. 
The present approach would permit
large-scale fluctuations in the transmission function at energies above
the fission barrier.  Pronounced peaks have been seen in $^236$U fission
above the barrier, but there they can also be ascribed to statistical 
fluctuations.

\end{abstract}

\maketitle

The theory of induced nuclear fission began 75 years ago
with Bohr and Wheeler's
landmark paper~\cite[Eq. (31)]{BW}, introducing the
transition-channel
formula for the fission decay rate $W$,
\be
\label{W}
W = { 1\over 2 \pi \hbar \rho_I} \sum_c T_c.
\ee
Here $\rho_I$ is the level density of the compound nucleus.  The $T_c$
are the transmission coefficients of the channels and satisfy the
condition $0< T_c < 1$.  The unit bound on the single-channel transmission
coefficient is an important aspect of the theory, derived from detailed 
balance.  In the Bohr-Wheeler theory the channel
concept is applied at the barrier top which is far from the asymptotic
region where the channels can be rigorously derived.  There is an
alternate formalism, the R-matrix theory, which forms the scaffolding
of present-day phenomenological parameterization of reactions leading
to fission~\cite{bj80,bo13,bo14}.  This theory is also based on the channel concept,
but there are no computational tools to calculate its basic 
ingredients such as the logarithmic derivative of the wave function
\cite[cf. p. 760]{bj80}.
The hallmark of well-developed channel physics
is the staircase excitation function, increasing by one step as each
new channel opens up.  This is by now
a well-known feature of quantum conductance, eg., see Ref. \cite{va88},
but conditions at the nuclear fission barrier are such
such as to obscure it from being visible in the excitation function.  Finally,
we mention that there is a new appreciation of importance of diffusive dynamics in nuclear
fission~\cite{ra11}; here channels play no role at all.

Besides the conductance through channels, there is another well-known
limit of electron transport in which the electrons 
pass from one conductor to another through an intermediate resonance, which
we shall call a ``bridge state". . The
formula for conductance is equivalent to the Bohr-Wheeler formula~\cite{be91}
but with a 
transmission coefficient $T_b$  given by the Breit-Wigner 
resonance expression~\cite{al00}
\be
T_b  =  {\Gamma_{R} \Gamma_{L} \over E_b^2 + (\Gamma_{R}+\Gamma_{L})^2/4}.
\ee 
Here $\Gamma_{R},\Gamma_{L}$ are the decays widths of the bridge state
into the two conductors and $E_b$ is its energy with respect
to the Fermi level of the electrons in the conductors.  
The main object
of the present study is to show 
how this structure arises in nuclear fission
via a discrete-basis representation of the many-body Hamiltonian.

A very simplified picture of the energy landscape of a fissile nucleus
is shown in Fig. 1.  The ground state is somewhat deformed, but to
fission the nucleus undergo a large shape changes that passes through
or over a barrier.  The colored regions in the left-hand graph show
different conditions that need to be considered for a complete theory:
the subbarrier tunneling region in green, the barrier-top region 
(most
relevant to fission in nuclear power reactors) in
blue, and the excited thermal region in red.  The barrier-top region is
most relevant to fission conditions in nuclear power reactions and is
the focus of the present work.  The dynamics for this region will be modeled
using a discrete basis of configurations characterized by a shape variable as
well as their energy.   

In fact discrete-basis representations arise naturally in the theory
of heavy nucleus structure.  
The most practical computational tool for that purpose is self-consistent
mean field theory (SCMF) based on energy density functionals~\cite{be03}, and 
simplified approximations thereto.  With the help of constraining
fields, the theory yields a spectrum of configurations characterized by
their energies and the expectation values of the constraining fields.
Thus it naturally produces a spectrum for a fissionable nucleus such as
that depicted on right-hand side of Fig. 1.  
\begin{figure}[tb]
\includegraphics[width=8cm]{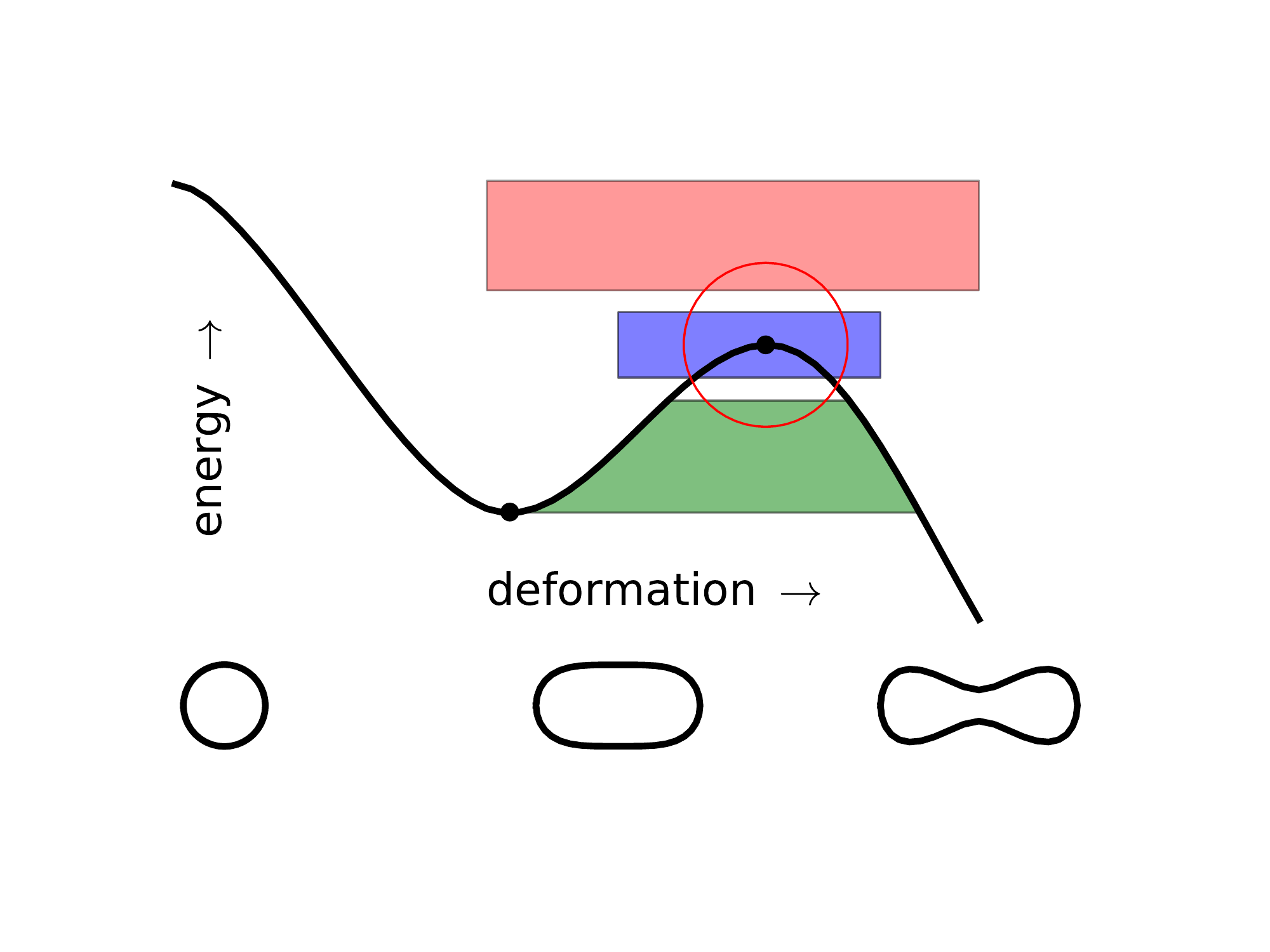}
\includegraphics[width=8cm]{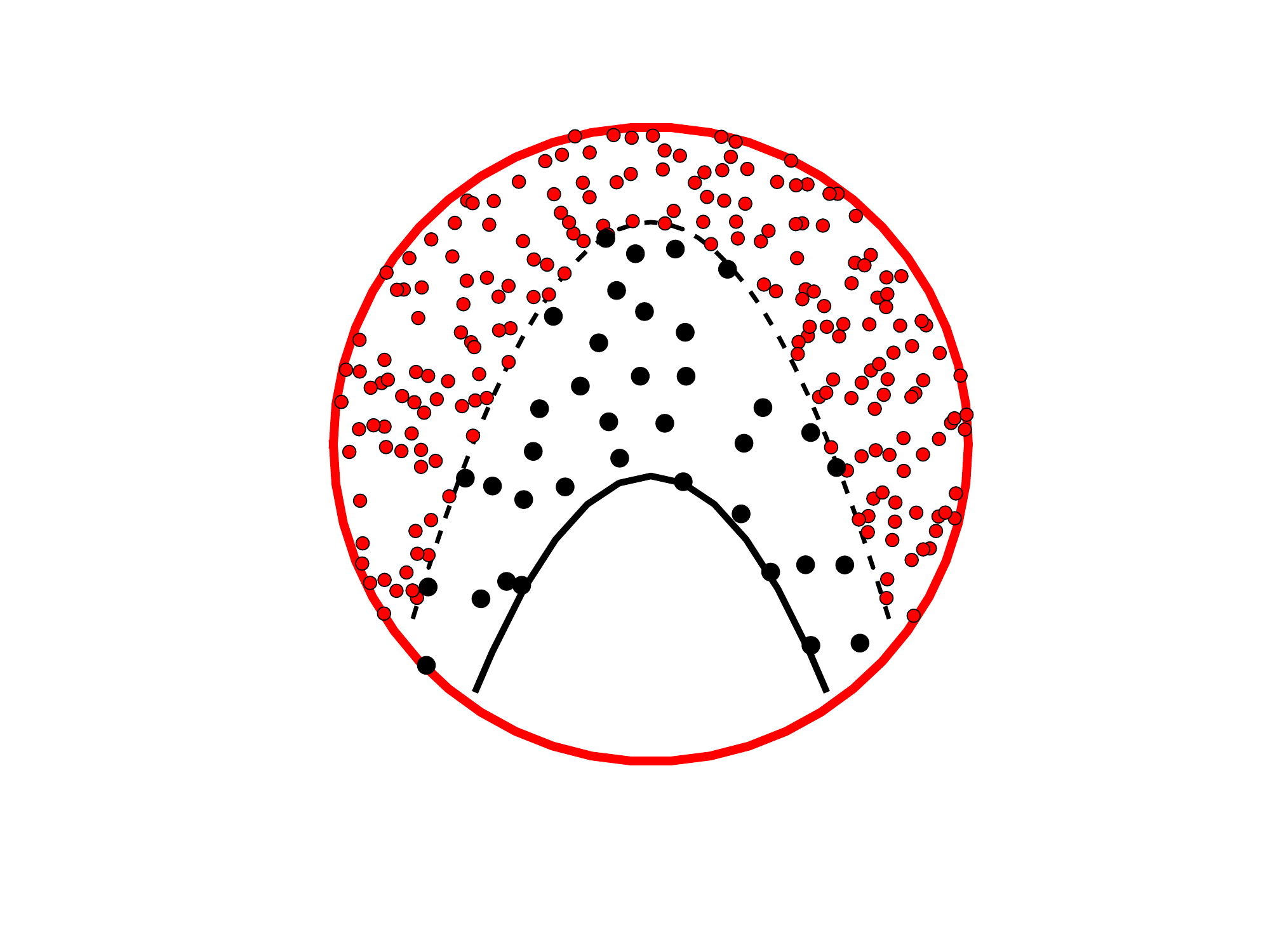}
\caption{\label{config-barrier} The fission barrier.  Top: the
black line depicts the potential energy surface (PES) as
a function of some deformation parameter associated with the 
shape change during fission.  The left-most black dot on the PES
indicates the moderately deformed ground state.
The other black 
dot on the PES indicates the barrier top.  The shaded areas show
regions of  different shape dynamics.  In the ground state and
at low excitation energy, fission occurs by tunneling through the barrier
(green area).  At sufficiently high energy, diffusive
dynamics is expected (red area).  The region just around the
barrier top (blue area) shows strong fluctuations and is the subject of the present
study.
Bottom:  a magnified view of the circled barrier-top region in
a discrete-basis representation.  Configurations are indicated by dots.
The solid black line is the PES near
the barrier top.  We distinguish regions
in the energy-deformation plane with low and high level densities. 
The high level-density regions are modeled by random matrix theory; the
the levels in the low-density region at the barrier must be treated
explicitly.
}
\end{figure}

There is also a residual interaction between the configurations that
is responsible for the dynamics.  The important point for the discrete
basis representation 
is that
that a two-body residual interaction cannot change the shape by a
a large amount, so the couplings are of limited range on the horizontal
scale.

At excitation energies relevant for the fission dynamics in
nuclear reactors, the configurations at the initial deformation form
a compound nucleus.  This means that the residual interactions produce
a statistical distribution of amplitudes and energies as in random
matrix theory.  The statistical limit is approached when the
residual interactions are uncorrelated and on average larger then 
the energy spacing between the levels~\cite{zi83}.  While this limit
is achieved on the right and left-hand sides at moderate excitation
energy, it will not be the case for
configurations in the middle, at energies around the barrier energy.
 
We thus a led to a Hamiltonian that consists of
of two statistical reservoirs
connected by a set of bridge configurations.  This already simplifies
the dynamics greatly.  One first determines the boundaries of the
reservoirs by calculation the level densities and average interaction
energies as a function of energy and the shape parameter.  The interactions
of the bridge configurations still has to be determined.  However, the
coupling to the reservoirs can be treated statistically:  the only
relevant parameter is the decay width into the reservoir.  

For the
remainder of this letter, and in order to make contact with
the Eq. (1) and (2),
we simplify the Hamiltonian to a single bridge state together with the
the reservoirs.  The time-dependent Schr\"odinger equation for the
Hamiltonian will be solved numerically to find the transition rate
out of the left-hand reservoir
\cite{supp}.

To define the Hamiltonian, we label the configurations in the left and 
right reservoirs by $l$ and $r$, 
respectively.  The bridge configuration is labeled by $b$.
The non-zero matrix elements of the Hamiltonian are:
$
\langle l | H | l' \rangle = \delta_{l,l'} E_l
$,
$
\langle r | H | r' \rangle = \delta_{r,r'} E_r
$,
$
\langle l | H | b \rangle = v_{lb}
$,
$
\langle r | H | b \rangle = v_{rb}
$,  and
$
\langle b | H | b \rangle = E_b.
$.

We first demonstrate that 
Eq. (2) of the mesoscopic
theory can be recovered by taking equally spaced levels in the 
reservoirs and constant values for the coupling matrix elements.
We denote the parameters as $v_L=v_{lb}, v_R=v_{rb},
\Delta E_L = E_l - E_{l-1}$
and $\Delta E_R = E_r- E_{r-1}$. 
It is convenient to define decay widths of the bridge state to the
right and left by Fermi's Golden Rule, 
$\Gamma_{L,R} = 2 \pi v^2_{L,R}/\Delta E_{L,R}$. 
As an example, we show in Fig. 2 (solid line) the survival probability
in the left-hand reservoir starting from an initial wave function
that is an eigenstate in the middle
of the spectrum of that reservoir.   The
Hamiltonian parameters are given in Figure caption.
\begin{figure}[tb]
\includegraphics[width=8 cm]{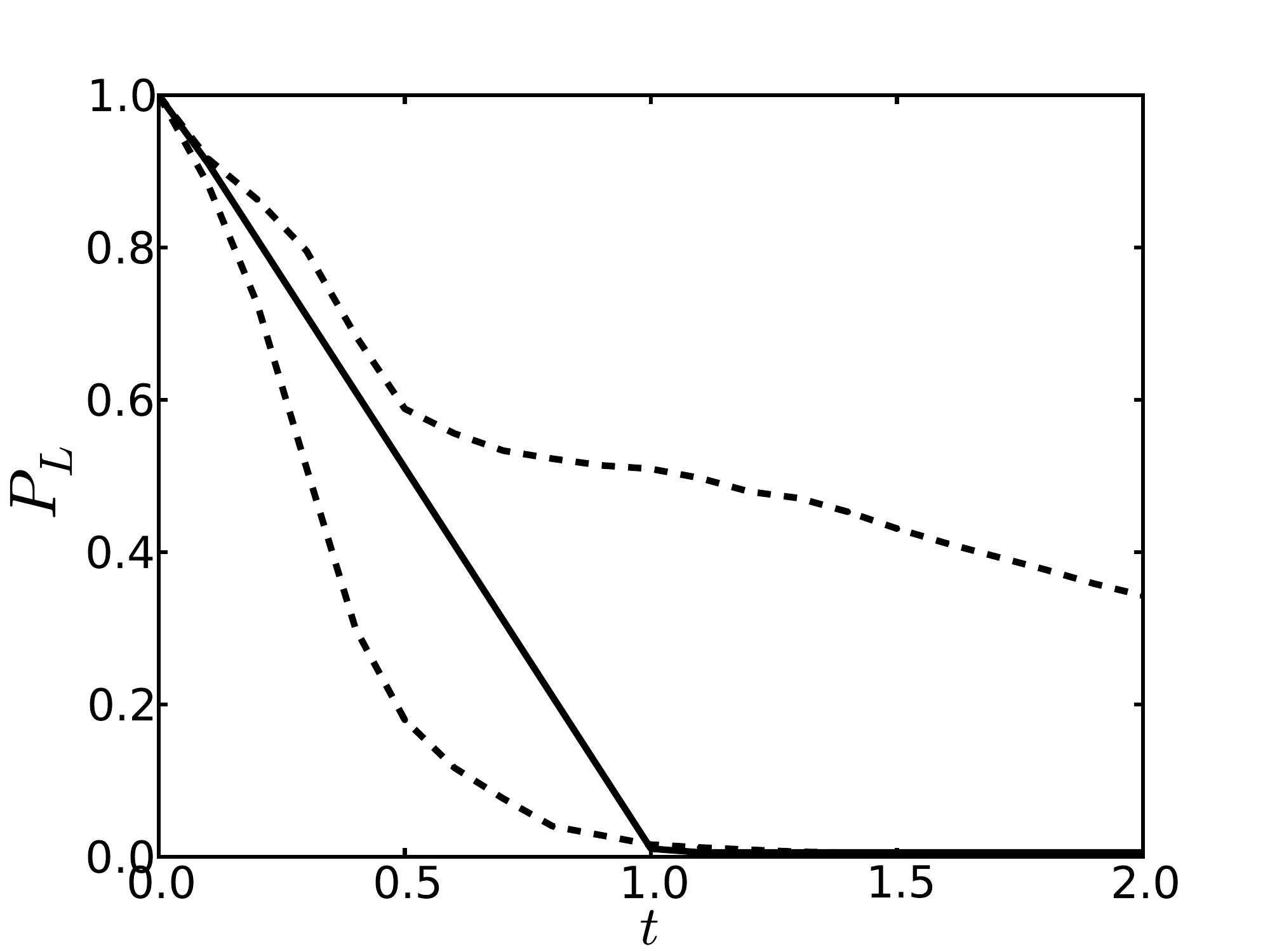}
\caption{\label{single-state-decay} Survival probability in the left-hand
reservoir as a function of time.
Solid black line:  regular Hamiltonians in 
both reservoirs. The numerical Hamiltonian has 100 and 200 levels in the 
left- and right-hand reservoirs respectively with the same bandwidths.
The coupling matrix elements are chosen so that $\Gamma_L = \Gamma_R$
and $\Gamma_L/\Delta E_L = 0.12 N_L$.  The energy of the bridge state
is taken equal to the initial state.  
Dashed lines:  two examples of decays when diagonal energies
of the left-hand Hamiltonian are taken from random matrix theory.  The
time $t$ is given in units of $t_0$, Eq. (\ref{t0}).
}
\end{figure}
One sees that the survival probability decreases linearly with time at
a rate $W = \Delta E_L /2 \pi$ up to the characteristic time
\be
\label{t0}
t_0 = {2 \pi \over \Delta E_L}.
\ee
This is in perfect agreement with the combined Eq. (1) and (2), since $T=1$ for
the given parameters.  It may seem surprising that the probability current
is constant up to the time $t_0$, but this can be easily understood.  The
uniform spacing of the levels in the left-hand reservoir simulates the middle
of a band in a perfect one-dimensional conducting wire.  The wave function
of an eigenstate of the isolated wire has the particle uniformly
distributed over 
the length of the wire, and equal currents flowing to the left and to the
right.  When the interaction with the bridge state is turned on, the
right-moving current passes without impediment to the other reservoir.  The
current only goes to zero after twice the transit time of the
wire.  If the parameters are changed so that $T<1$, the only difference up
to a time $t_0$ is that the slope changes from $t_0^{-1}$ to $T/t_0$.

We now go to Hamiltonians closer to the nuclear cases, modifying the
reservoirs according to random matrix theory.   In fact, we found that
the physics associated with the right-hand reservoir is 
insensitive to the fine details
of its Hamiltonian.  The levels spacings can be taken to be uniform or 
as extracted from the middle levels from Wigner's random matrix ensemble.
The coupling matrix elements $v_{rb}$ can be constant or Gaussian 
distributed, again from Wigner's random matrix ensemble.  The only important
properties are:
\begin{quotation}
\noindent  
$\bullet$ The width $\Gamma_R$.  It may be computed from the 
Fermi Golden Rule using average level density and the
root-mean-square average interaction matrix element.\\
$\bullet$ The average level spacing must be smaller than any other
energy scale (or inverse time scale).
\end{quotation}
Under these conditions the coupling to the right-hand reservoir can be
treated very simply.  Instead of including it explicitly in the 
Hamiltonian, the absorptive effect of the right-hand reservoir can be
computed adding an imaginary potential $-i \Gamma_R/2$ to $\langle b | H |
b \rangle$.  We have adopted this simplification for computing the 
rates shown below.

In sharp contrast to the results for uniform level
spacing in the left-hand reservoir, the decay properties 
are quite different with reservoir governed by random matrix theory.
One aspect is well-understood:  the
decay width of a state $l$ is proportional to the square of the coupling
matrix element $v_{lb}$ which has a Gaussian distribution in random
matrix models.  This is called the Porter-Thomas distribution.  But there
is more.  Fig. 2 shows two decay distributions when the level spectrum
$E_l$ was taken from a random matrix ensemble but with constant coupling
matrix elements, as in the quantum wire.
One sees that the decay rates do not have
any simple behavior; they can neither be described as constant or
as exponential decays.

For the physical problem we only care about the averages over
many initial states.  Fig. 3 shows such an average, for conditions
that correspond to a unit transmission coefficient, $\Gamma_L =
\Gamma_R$ and $E = E_b$.  
\begin{figure}[tb]
\includegraphics[width= 10cm]{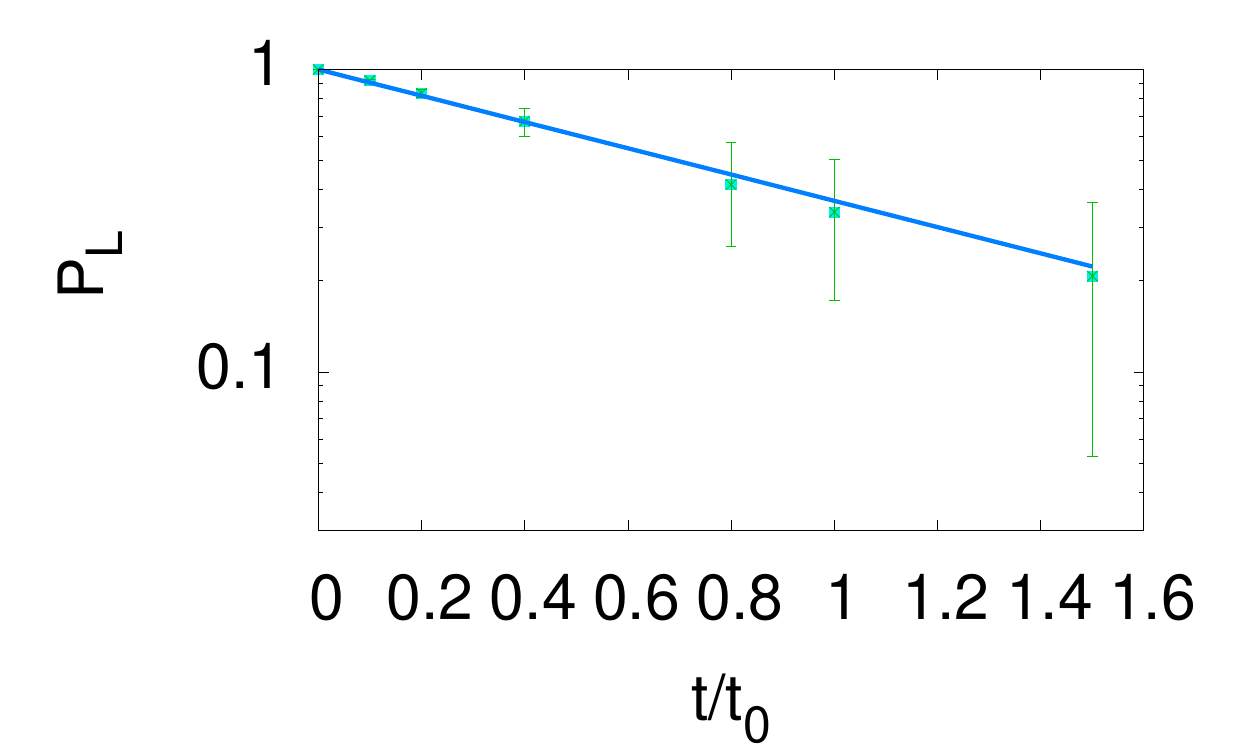}
\caption{\label{decay-average1} Survival probability in the the left-hand
reservoir averaged over the random statistical ensemble
of reservoir Hamiltonians but with a constant coupling to the bridge
state.  Squares show the average values and
error bars show the 
the root-mean-square deviation of individual decay probabilities.  Solid
line is the exponential function $\exp(-W t)$ with $W$ from Eq.
(1).
}
\end{figure}
The average follows very well an 
exponential decay law, with an average decay rate given by Eq. (1).
The error bars show the root mean square deviation of the individual
probabilities $P(t)$; one sees that there are large fluctuations about the
average.  We have also 
examined the dependence of the decay profiles $ P (t)$ on 
$\Delta E_L$, $v_L$, and $E_b$ and found that the average profiles
are exponential and fairly well described by the formula
\be\label{gamma}
\overline \Gamma_c = {1\over 2 \pi \rho_I} 
 {\Gamma_R \Gamma_L \over E_b^2 + (\Gamma_R+\Gamma_L)^2/4}.
\ee

As a final step to apply random matrix distributions to the left-hand reservoir,
we taking the coupling strengths $v_{bL}$ to be Gaussian distributed with
variance $\langle v^2_{bL} \rangle = v_0^2$.  
The expected average survival probability is then given
by  
\be\label{full}
\overline P(t) = 
{1\over (2\pi v_0)^{1/2}}
\int_{-\infty}^\infty d v  
e^{-v^2/2v_0^2 - 2 \pi v^2 \rho_I t}
={1\over (1 + 2 \Gamma_0 t)^{1/2}},
\ee    
where $\Gamma_0 = 2 \pi v_0^2 \rho_I$.
Fig. 4 shows the computed $\overline P(t)$ as black dots, taking parameters
such than $T_b = 1$.  It agrees very well with
Eq. (\ref{full}), shown as the solid line in the Figure.  For these
parameters, $\Gamma_0= 1/2 \pi \rho_I= t_0^{-1}$.  
\begin{figure}[tb]
\includegraphics[width= 9cm]{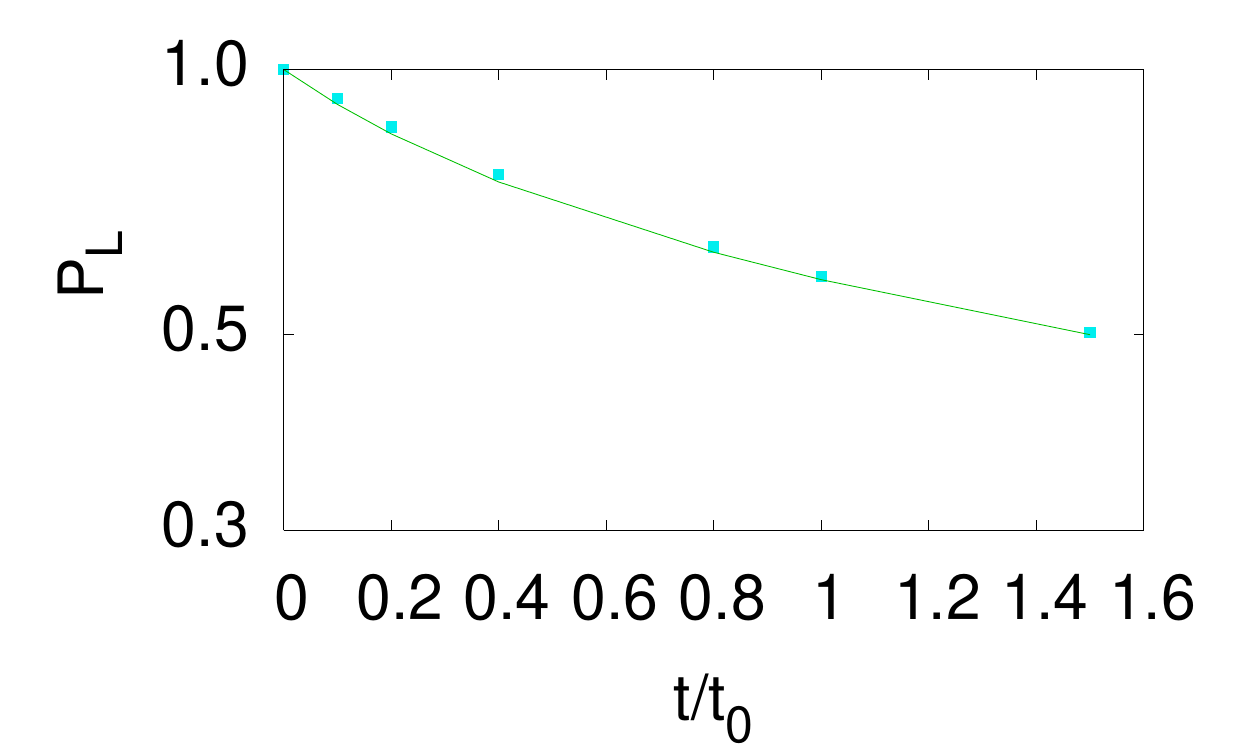}
\caption{\label{decay-average2} Squares show the survival probability in the left-hand
reservoir averaged over the random ensemble of Hamiltonians, including
the random distribution of couplings to the bridge state.
Eq. (\ref{full}) is shown as the solid line.
}
\end{figure}

A hallmark of the present approach would be the observation of fluctuations
in the fission transmission function above the barrier that could be
ascribed to individual bridge states.  This would require that the 
density of bridge states not be too large compared to the decay widths
of the states.  Detailed calculations of the coupling matrix elements 
may be required to see if that is likely for the lowest states.  

Experimentally \cite{mo78}, large-amplitude fluctuations have been seen in the
suprabarrier fission of $^{236}$U on the scale of 1 keV .  The amplitude
of the fluctuation in $T$ is of the order 1, which is consistent with a
resonant mechanism.  However, the spacing of bridge states would be much
larger than 1 keV, since the excitation energies of the observed
fluctuations is only 1 MeV above the barrier.  In fact, these fluctuations
can be accounted for as statistical in origin, arising from the
Porter-Thomas fluctuations of decays of the intermediate states in the
fission process \cite{mo84}.

The discussion in the above paragraph hints at the many complications that
are present in making a more realistic description of fission.
We mention two
of the important ingredients for a predictive theory that have been
neglected here.  The first is that the barrier region very likely requires
many bridge configurations to be considered explicitly.  The fission barrier
has at least two humps~\cite{br72}.  Besides bridge states crossing the
two humps, the intermediate states, called class II states, are visible 
as closely-spaced resonances in the fission excitation functions.  Once we go beyond the
individual bridge state linking the two reservoirs, the properties of the
interaction linking the bridge states becomes very important.  It is
well-known that in the sub-barrier region the fission lifetimes are very
sensitive to the pairing interaction \cite{ro14}.  This gives a coherence
to the matrix elements of the lowest bridge states, and if the pairing
strength were large enough would allow the linked states to act as channels.
However, nuclear pairing is rather weak and is easy blocked in excited 
states.  For that reason, the discrete basis picture may be closer to the
physical reality than the channel picture.

Acknowledgments.  This work arose out of discussions in the program
``Quantitative large amplitude dynamics" at the Institute
for Nuclear Theory, with especial thanks to W. Nazarewicz for encouraging
this line of inquiry.  The author also thanks Y. Alhassid and 
R. Vandenbosch
for helpful discussions.  Finally, the author thanks P. Talou and O.
Bouland for pointing out the importance of the statistical mechanism to
introduce fluctuations in the suprabarrier transmission function.
Research was support by the DOE under Grant No. FG02-00ER41132.

\end{document}